\documentclass[a4paper,12pt]{article}
\usepackage[german]{babel}
\usepackage[utf8x]{inputenc}
\usepackage{amsmath,amssymb,amsthm}
\usepackage[margin=3cm]{geometry}
\usepackage{enumerate}
\usepackage{hyperref}
\usepackage{graphicx}
\title{Relativistische Materie in zwei Raum-Zeit-Dimensionen}

\author{Henning Bostelmann}

\date{Vortrag im Symposium ,,Raum und Materie``,\\ Haus Villigst, 15.-18.~Oktober 2012}

\usepackage{amsmath,amsfonts}
\usepackage{latexsym}

\def \cbb{\mathbb{C}}

\def \rbb{\mathbb{R}}

\def \ocal {\mathcal{O}}

\def \wcal {\mathcal{W}}

\newcommand{\thetav}{\boldsymbol{\theta}}
\newcommand{\etav}{\boldsymbol{\eta}}
\newcommand{\zetav}{\pmb{\zeta}}

\def \<  {\langle}
\def \>  {\rangle}

\def \. { \,\! }

\DeclareMathOperator{\im}{Im}

\newcommand{\idop}{\boldsymbol{1}}

\newcommand{\ad}{a^{\dagger}}
\newcommand{\zd}{z^{\dagger}}

\newcommand{\A}{\mathfrak{A}}

\def \st {^\ast}

\DeclareMathOperator{\supp}{supp}

\def \expltext#1 {\\ \text{\footnotesize{ (#1) }}\\}
\def \intercomm#1 {\\ \text{\footnotesize{ (#1) }}\\}
\def \undercomm#1 {\underset{\text{\scriptsize{ (#1) }}}}
\def \overcomm#1 {\overset{\text{\scriptsize{ (#1) }}}}

\newcommand{\Hil}{\mathcal{H}}

\def \tracenorm#1 { \| #1 \| _1 }
\def \hsnorm#1 { \| #1 \| _2 }

\def \vectorcomp#1 {
  \left( \begin{array}{c}
  #1
  \end{array}
  \right)  }

\def \pder#1#2 { \frac{ \partial #1 }{ \partial #2 }}

\begin{document}

\maketitle

\begin{abstract}
Quantenfeldtheorie vereinigt die Konzepte der Quantentheorie mit denen der speziellen Relativitätstheorie. Ihre mathematisch strenge Beschreibung ist aufwändig und nur teilweise verstanden; das gilt insbesondere für die Konstruktion von Operatoren, die Messungen an einem Raum-Zeit-Punkt oder in einem beschränkten Raum-Zeit-Gebiet beschreiben. Wir erläutern dies anhand vereinfachter Modelle in 1+1-dimensionaler Raumzeit, und zwar sogenannter integrabler Modelle. Wir geben eine Charakterisierung lokaler Operatoren durch die Analytizitätseigenschaften ihrer Koeffizienten in einer Reihenentwicklung an. Dies erlaubt auch die explizite Konstruktion von Beispielen lokaler Operatoren.
\end{abstract}

\section{Einleitung}

Die Quantentheorie und die spezielle Relativitätstheorie sind beide seit mehr als 100 Jahren bekannt; dennoch ist es nach wie vor schwierig, ihre Grundlagen miteinander zu vereinbaren. Zwar ist die Berücksichtigung relativistischer Korrekturen in der Atomphysik oder der Quantenchemie durchaus gängig; und in der Hochenergiephysik liefert die störungstheoretische Quantenfeldtheorie Vorhersagen, die gut mit dem Experiment übereinstimmen. Nach wie vor ist die mathematische Formulierung dieser Systeme aber unbefriedigend: Die üblichen Methoden der Störungstheorie liefern lediglich formale Potenzreihen, deren Konvergenz meist nicht kontrolliert werden kann. 

Ziel der \emph{axiomatischen Quantenfeldtheorie} ist es, relativistische quantentheoretische Systeme ohne diese Divergenzen zu beschreiben. Dies gelingt bisher nicht vollständig, jedoch sind wichtige Teilresultate erzielt worden. Zunächst ist es möglich, mathematische Axiome aufzustellen, die beschreiben, welche Eigenschaften ein mathematisches Modell für die Hochenergiephysik haben sollte, z.B.~im Sinne der Wightman-Axiome \cite{StrWig:PCT} oder der Haag-Kastler-Axiome \cite{Haa:LQP}. Aus diesen Axiomen -- die für \emph{alle} quantenfeldtheoretischen Modelle gültig sein sollen, die also z.B.~das Teilchenspektrum oder das Wechselwirkungspotential nicht festlegen -- lassen sich dann verschiedene Konsequenzen herleiten, wie etwa eine Teilcheninterpretation im Sinne von Streutheorie \cite{BuchholzSummers:2006} oder die Existenz thermodynamischer Gleichgewichtszustände \cite{BucJun:equilibrum}. Schwierigkeiten bestehen jedoch nach wie vor dabei, konkrete Modelle zu konstruieren, die diesen Axiomen genügen. Insbesondere ist nach wie vor kein solches wechselwirkendes Modell in physikalischen Raum-Zeit-Dimensionen konstruiert worden. (Nicht ohne Grund ist die 
Konstruktion von Yang-Mills-Theorie in physikalischer Raumzeit eines der Probleme des ,,Millennium Prize`` und mit 1.000.000 US-Dollar Preisgeld dotiert. \cite{JaffeWitten:millenium})

Um dennoch die Konsequenzen der Axiome studieren zu können, verwendet man oft vereinfachte ,,Spielzeugmodelle`` in niedrigdimensionaler Raum-Zeit, insbesondere in 1+1 Dimensionen. Hier ist die mathematische Beschreibung von Wechselwirkung weit weniger komplex. Ein bekanntes Beispiel sind die von Glimm und Jaffe konstruierten $P(\phi)_2$-Modelle \cite{GliJaf:quantum_physics}. 

Wir werden hier eine andere Klasse 1+1-dimensionaler Quantenfeldtheorien betrachten, die so genannten \emph{integrablen Modelle}, in denen das Streuverhalten der Teilchen stark vereinfacht ist; insbesondere werden durch die Wechselwirkung keine Teilchen erzeugt oder vernichtet. Ein Beispiel dafür ist die sinh-Gordon-Theorie. Im Folgenden werden wir den Aufbau dieser Modelle beschreiben und insbesondere die Struktur von \emph{lokalen Observablen} untersuchen, d.h. physikalisch von Messungen, die in einem endlichen Zeitintervall und in einem begrenzten Raumgebiet ausgeführt werden.

Die axiomatische Quantenfeldtheorie ist ein mathematisch sehr anspruchsvolles Gebiet und verlangt den Einsatz fortgeschrittener Methoden z.B.~der Funktionalanalysis. Viele dieser mathematischen Details werde ich in diesem kurzen \"Uberblick jedoch überspringen oder sie nur kurz für Leser mit entsprechendem Hintergrundwissen erwähnen. Details über das vorgestellte Thema finden sich in \cite{Lechner:2008,Cadamuro:2012,BostelmannCadamuro:expansion}.
 
\section[Wechselwirkung in zwei Dimensionen]{Wechselwirkung in zwei Dimensionen -- von der klassischen Mechanik zur Quantenfeldtheorie}

\begin{figure}
\begin{center}
\begin{minipage}{0.45\textwidth}
\begin{center}
 \includegraphics[width=\textwidth]{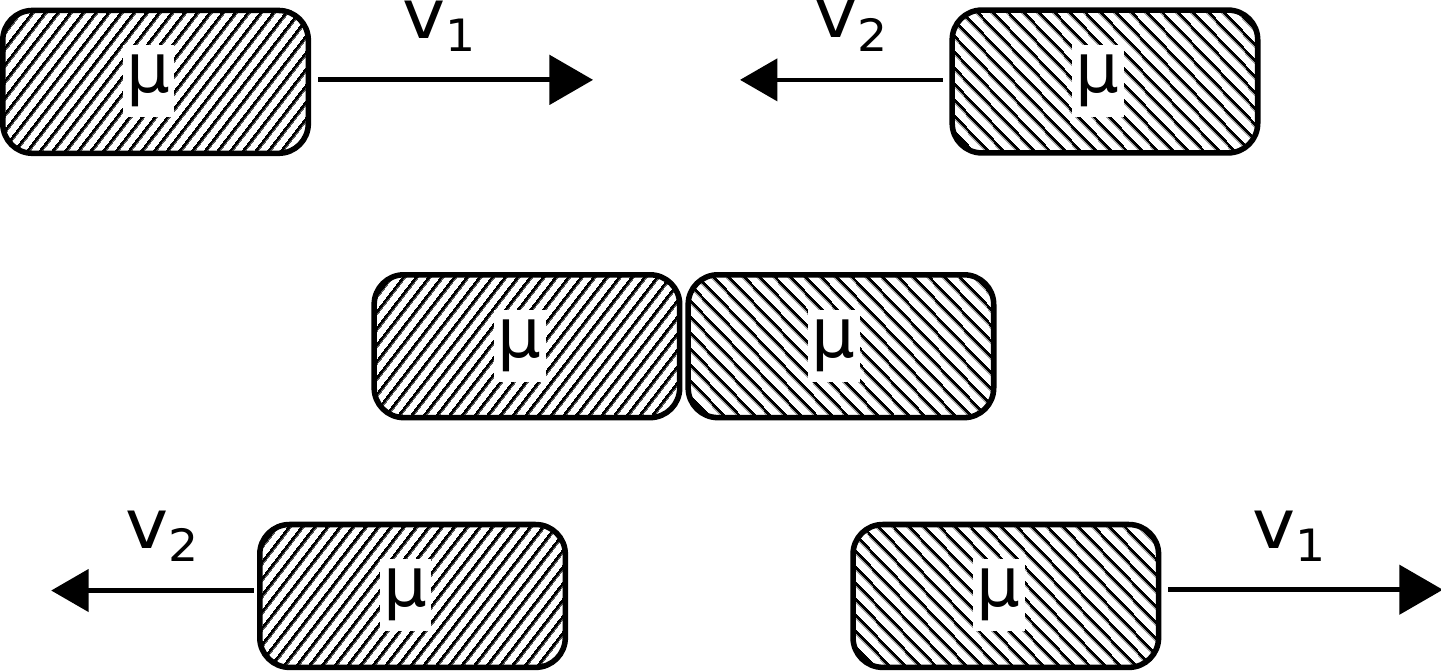}
 \caption{Elastischer Stoß zweier Massenpunkte mit gleicher Masse $\mu$}\label{fig:collision}
 \end{center}
\end{minipage}
\quad
\begin{minipage}{0.45\textwidth}
\begin{center}
 \includegraphics[width=0.95\textwidth]{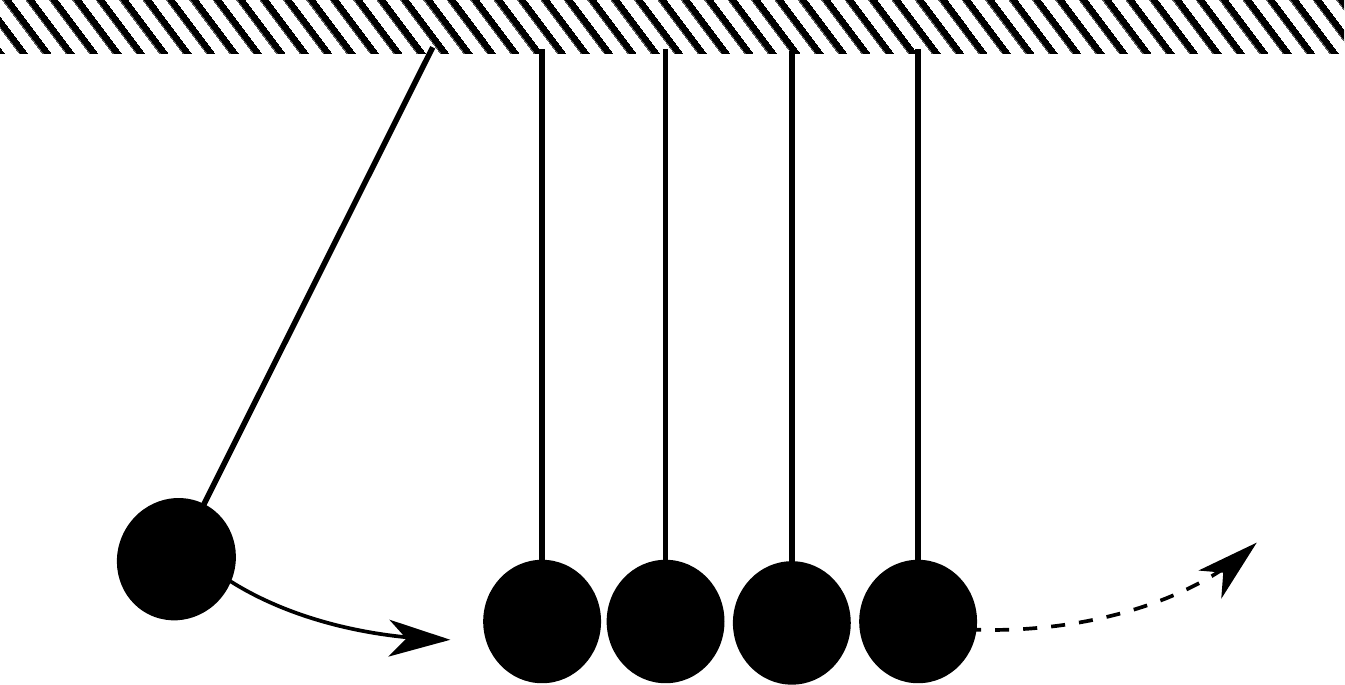}
 \caption{Kugelstoßpendel (,,Newton's \mbox{cradle}``)}\label{fig:cradle}
 \end{center}
\end{minipage}
\end{center}
\end{figure}

Um die Art der Wechselwirkung in integrablen Modellen zu verstehen, beschreiben wir zunächst deren Analogon in der klassischen Mechanik. Wir betrachten Punktmassen mit gleicher Masse $\mu$, die sich in einer räumlichen Dimension bewegen -- etwa Stahlkugeln auf einer Stange oder an einem Pendel. Als Wechselwirkung zwischen zwei Massen lassen wir nur Kontaktwechselwirkung zu, d.h., einen elastischen Stoß. Der Energie- und der Impulserhaltungssatz legen das Ergebnis dieses ,,Streuexperiments`` dann eindeutig fest: Die zwei am Stoß beteiligten Kugeln tauschen einfach ihre Geschwindigkeiten\footnote{Wenn wir die Unterscheidbarkeit der Massen aufgeben würden -- wie später in der Quantentheorie -- dann sähe die Wechselwirkung also genauso aus wie der wechselwirkungsfreie Fall, d.h., wenn die Massen sich ohne Berührung aneinander vorbei bewegen.}
(Abb.~\ref{fig:collision}). 

Für die Wechselwirkung zwischen mehr als zwei Teilchen wird man normalerweise annehmen, dass sie aus Folgen von Zweiteilchen-Stößen besteht. Dies ist das Prinzip des Kugelstoßpendels (\emph{Newton's cradle,} Abb.~\ref{fig:cradle}).

Wir übertragen dasselbe Prinzip nun in die spezielle Relativitätstheorie, zunächst noch ohne Berücksichtigung der Quantentheorie. Wir betrachten also Punktmassen mit Masse $\mu>0$ nahe der Lichtgeschwindigkeit. In diesem Fall werden Energie und räumlicher Impuls eines Teilchens zu einem Zweier-Impuls $p = (p^0,p^1)$ zusammengefasst, wobei $p_\alpha p^\alpha  = \mu^2$, $p^0>0$ gilt.\footnote{%
Setzt man hier $p^0=E$, $p^1=0$, und schreibt $m$ statt $\mu$ für die Teilchenmasse, so ergibt sich die bekannte Relation $E=mc^2$ -- wir verwenden Einheiten, in denen $c=1$ gilt.}
Solche Zweier-Impulse lassen sich am besten durch ihre \emph{Rapidität} $\theta$ parametrisieren, und zwar gilt
\begin{equation}
    p=p(\theta) = \mu (\cosh \theta, \sinh \theta), \quad \theta \in \rbb.
\end{equation}
Die Streutheorie in diesem relativistischen Modell ist sehr ähnlich zur nichtrelativistischen Mechanik: Sind zwei Teilchen (A und B) am Stoß beteiligt, so gelten zwei Erhaltungssätze, nämlich die beiden Komponenten des Gesamtimpulses sind erhalten: $p_{(A)}+p_{(B)}=\text{const.}$ Das Ergebnis entspricht genau dem nichtrelativistischen Verhalten: Die zwei Teilchen ,,tauschen`` ihre Rapiditäten. Nach wie vor nehmen wir an, dass Mehrteilchenstreuung durch Verkettung von Zweiteilchen-Streuprozessen beschrieben wird.

Wir beschreiben nun kurz den \"Ubergang zur relativistischen \emph{Quantentheorie}. Analog zu den klassischen Massenpunkten betrachten wir dort Bosonen mit Spin 0 und Masse $\mu>0$, die sich in einer 1+1-dimensionalen Raum-Zeit bewegen. Die Streutheorie zwischen zwei Teilchen ist in diesem Fall etwas reichhaltiger als im klassischen Fall: Zwar gilt nach wie vor die Erhaltung des Zweier-Impulses bei Streuung, jedoch ist der Austausch eines \emph{Phasenfaktors} bei Streuung möglich. Aus Gründen von Lorentz-Kovarianz kann diese Phase nur von der Differenz der Teilchen-Rapiditäten abhängen. Die Zweiteilchen-Streumatrix ist daher ein Phasenfaktor $S(\theta_A-\theta_B)$. Diese Funktion $S$ soll dabei analytisch im Streifen $0 < \im \theta <\pi$ sein und muss gewisse Symmetrieeigenschaften erfüllen, auf die wir hier nicht näher eingehen. Analog zur klassischen Mechanik wollen wir ein Modell konstruieren, in der die Streuung mehrerer Teilchen als Abfolge von Zweiteilchen-Streuung verstanden werden kann; man spricht dann von einem Modell mit \emph{faktorisierender Streumatrix}. Wichtige Spezialfälle sind die zwei Fälle konstanter Zweiteilchen-Streufunktion, $S=1$ (dies entspricht nicht-wechselwirkenden Teilchen) und $S=-1$ (das so genannte \emph{Ising-Modell}). Ein Beispiel für nicht-konstantes $S$ liefert das sinh-Gordon-Modell:
\begin{equation}
  S(\theta) = \frac{\sinh \theta - i \sin b}{\sinh \theta + i \sin b},
\end{equation}
wobei $0<b<\pi$ eine modifizierte Kopplungskonstante ist.

Die Aufgabe besteht nun darin, zu gegebenem $S$ eine mathematisch konsistente Quantenfeldtheorie zu konstruieren, die der heuristischen Beschreibung oben entspricht. Man könnte dies als ,,inverse Streutheorie`` beschreiben, da diese Konstruktion die Funktion $S$ und nicht die Lagrangefunktion o.ä.~als Ausgangspunkt verwendet. Wir werden die Lösung im Sinne von \cite{Lechner:2008} nun kurz skizzieren.

\section{Mathematische Beschreibung des Modells}

Unsere Konstruktion integrabler Modelle ist in weiten Teilen analog zur freien Quantenfeldtheorie, die in gewissem Sinne ,,deformiert`` wird, um die wechselwirkende Situation zu erhalten. Wir erinnern daher zunächst an die Beschreibung des reellen skalaren \emph{freien} Feldes in zwei Dimensionen. Diese basiert auf Erzeugungs- und Vernichtungsoperatoren $\ad(\theta)$, $a(\theta)$, welche kanonische Vertauschungsrelationen erfüllen (CCR-Algebra):
\begin{equation}\label{eq:ccr}
\begin{aligned}
 a(\theta_1)  a(\theta_2)  &=  a(\theta_2)a(\theta_1)\,,\\
 \ad(\theta_1) \ad(\theta_2) &= \ad(\theta_2)\ad(\theta_1)\,,\\
 a(\theta_1)   \ad(\theta_2) &= \ad(\theta_2)a(\theta_1)+\delta(\theta_1-\theta_2)\cdot \idop.
\end{aligned}
\end{equation}
Als Operatoren wirken die $\ad(\theta)$, $a(\theta)$ auf dem Fockraum $\Hil$, der von allen $n$-Teilchenvektoren der Art
\begin{equation}
   \psi_n = \int d^n\theta \, f(\theta_1,\ldots,\theta_n)\, \ad(\theta_1)\ldots \ad(\theta_n)\Omega
\end{equation}
aufgespannt wird. Auf $\Hil$ sind die Symmetrien der Raumzeit (Translationen $T_x$, Lorentz-Boosts $B_\lambda$ und die Raumzeit-Spiegelung $j$) dargestellt durch
\begin{align}
  {U(T_x)} \ad(\theta_1)\ldots \ad(\theta_n)\Omega &= e^{i\sum_j p(\theta_j) \cdot x} \ad(\theta_1)\ldots \ad(\theta_n)\Omega , \label{eq:utx}
\\
  {U(B_\lambda)} \ad(\theta_1)\ldots \ad(\theta_n)\Omega  &= \ad(\theta_1+\lambda)\ldots \ad(\theta_n+\lambda)\Omega ,  \label{eq:ul}
\\
  {U(j)} \ad(\theta_1)\ldots \ad(\theta_n)\Omega &= \ad(\theta_1)\ldots \ad(\theta_n)\Omega,  \label{eq:uj}
\end{align}
wobei \eqref{eq:utx} und \eqref{eq:ul} linear, aber \eqref{eq:uj} \emph{antilinear} auf ganz $\Hil$ fortgesetzt werden.

Die wechselwirkende Theorie (für eine gegebene Streufunktion $S$) wird nun als eine Deformation der freien Theorie aufgebaut. Anstatt der CCR-Algebra betrachten wir ,,wechselwirkende`` Erzeuger und Vernichter $\zd(\theta)$, $z(\theta)$, die den Relationen der \emph{Zamolodchikov-Faddeev-Algebra} genügen:
\begin{equation}\label{eq:zfr}
\begin{aligned}
 z(\theta_1)    z(\theta_2)  &= {S(\theta_1-\theta_2)} \, z(\theta_2)z(\theta_1)\,,\\
 \zd(\theta_1) \zd(\theta_2) &= {S(\theta_1-\theta_2)} \,\zd(\theta_2)\zd(\theta_1)\,,\\
 z(\theta_1)   \zd(\theta_2) &= {S(\theta_2-\theta_1)} \,\zd(\theta_2)z(\theta_1)+\delta(\theta_1-\theta_2)\cdot \idop.
\end{aligned}
\end{equation}
Der ,,$S$-symmetrische`` Fockraum $\Hil$, auf dem $\zd(\theta)$ und $z(\theta)$ wirken, wird aufgespannt durch die $n$-Teilchen-Vektoren
\begin{equation}
   \psi_n = \int d^n\theta \, f(\theta_1,\ldots,\theta_n)\, \zd(\theta_1)\ldots \zd(\theta_n)\Omega,
\end{equation}
und die Raum-Zeit-Symmetrien sind dargestellt als
\begin{align}
  {U(T_x)} \zd(\theta_1)\ldots \zd(\theta_n)\Omega &= e^{i\sum_j p(\theta_j) \cdot x} \zd(\theta_1)\ldots \zd(\theta_n)\Omega ,
\\
  {U(B_\lambda)} \zd(\theta_1)\ldots \zd(\theta_n)\Omega  &= \zd(\theta_1+\lambda)\ldots \zd(\theta_n+\lambda)\Omega ,
\\
  {U(j)} \zd(\theta_1)\ldots \zd(\theta_n)\Omega &= \zd(\theta_n)\ldots \zd(\theta_1)\Omega,
\end{align}
wobei die letzte Relation wiederum antilinear fortgesetzt wird. 

\begin{figure}
\begin{center}
\begin{minipage}{0.45\textwidth}
\begin{center}
 \includegraphics[width=0.97\textwidth]{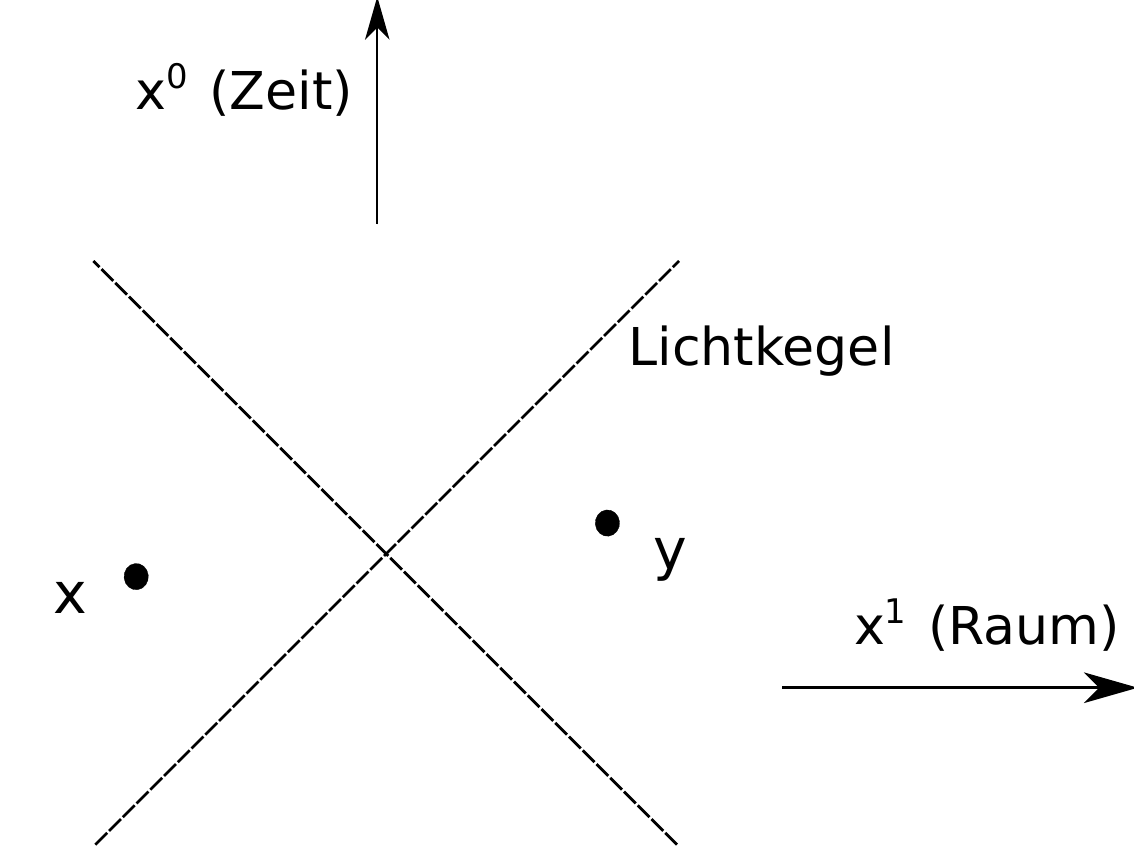}
 \caption{Raumartig getrennte Punkte $x$ und $y$ in zweidimensionaler Raum-Zeit}\label{fig:spacelike}
 \end{center}
\end{minipage}
\quad
\begin{minipage}{0.45\textwidth}
\begin{center}
 \includegraphics[width=\textwidth]{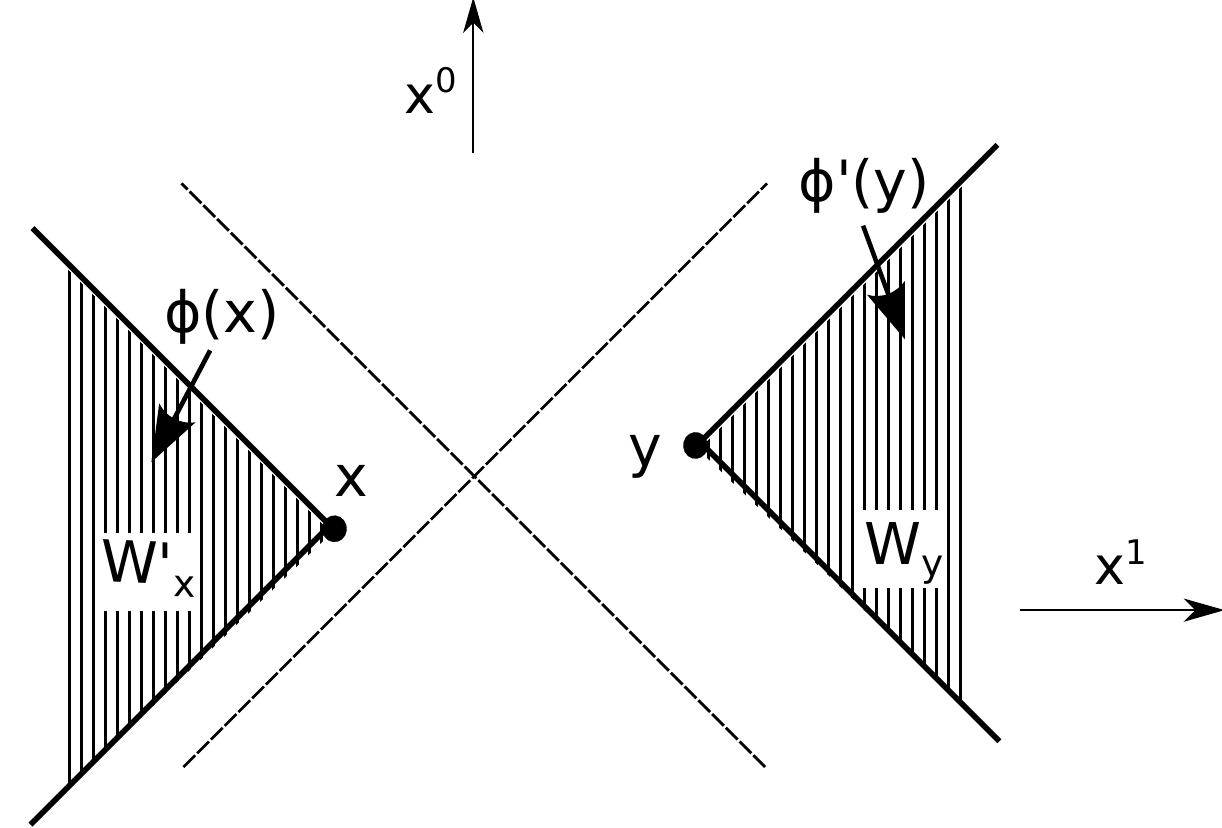}
 \caption{Die Lokalisationsgebiete der Felder $\phi(x)$ und $\phi'(y)$}\label{fig:wedges}
 \end{center}
\end{minipage}
\end{center}
\end{figure}

Von entscheidender Bedeutung ist nun die Konstruktion \emph{lokaler Observabler} des Modells, welche das relativistische Kausalitätsprinzip widerspiegeln. Wir erinnern dazu an den Begriff von \emph{raumartigen Abständen}: Ein Raum-Zeit-Punkt $x$ liegt raumartig zu einem anderen Punkt $y$ (in anderen Worten, $x$ und $y$ sind raumartig getrennt), wenn $x$ und $y$ so weit räumlich auseinander liegen, dass kein Lichtsignal -- und damit kein anderer kausaler Einfluss -- von $x$ nach $y$  und umgekehrt gelangen kann (Abb.~\ref{fig:spacelike}). Zwei physikalische Messungen, die lokal in $x$ bzw. $y$ stattfinden -- man mag hier an Messungen der Energiedichte bei $x$ bzw. $y$ denken --, müssen daher in gewissem Sinne unabhängig sein. In der Quantentheorie drückt sich das so aus, dass die zugehörigen Operatoren kommutieren. Etwas formaler heißt dies: Sind $A(x)$ und $B(y)$ zwei lokale Observable, abhängig von zwei Raum-Zeit-Punkten $x$ und $y$, dann soll gelten, dass
\begin{equation}
  [ A(x), B(y) ]  = 0  \quad \text{falls $x$ raumartig zu $y$ liegt.}
\end{equation}
Der Begriff ,,Operator`` ist hier etwas ungenau verwendet, denn in der Regel wird die obige Gleichung nur für mathematisch sehr singuläre Objekte (unbeschränkte quadratische Formen) erfüllt sein. Man kann dies vermeiden, indem man sich statt auf Raum-Zeit-Punkte $x$ auf ausgedehnte Raum-Zeit-Gebiete $\ocal$ bezieht. In der Quantenfeldtheorie möchte man dann für jedes Gebiet $\ocal$ eine Menge von Observablen $\A(\ocal)$  erhalten, so dass
\begin{equation}\label{eq:ABlocal}
  \begin{aligned}
      {\lbrack A, B \rbrack  = 0} \quad &{\text{falls $A \in \A(\ocal_1)$, $B \in \A(\ocal_2)$} }, \\ &
                   \text{und $\ocal_1$ raumartig zu $\ocal_2$ liegt.}
   \end{aligned}
\end{equation}
Man stellt an diese Zuordnung von Raum-Zeit-Gebieten zu Mengen von Operatoren, $\ocal \mapsto \A(\ocal)$, noch weitere Anforderungen, in der sich u.a.~Poincar\'e-Invarianz und Positivität der Energie ausdrücken \cite[Ch.~III]{Haa:LQP}; wir gehen darauf hier nicht näher ein. Die Mengen $\A(\ocal)$ wählt man normalerweise als Algebren beschränkter Operatoren ($C\st$-Algebren oder von-Neumann-Algebren).

Bevor wir uns nun den lokalen Observablen in unseren wechselwirkenden Modellen zuwenden, wiederholen wir kurz die Situation für das freie Quantenfeld in 1+1 Dimensionen. Wir definieren den Operator
\begin{equation}\label{eq:phidef}
  \phi(x) := \int d\theta \, \Big(e^{ip(\theta) \cdot x} \ad(\theta) + e^{-ip(\theta) \cdot x} a(\theta) \Big).
\end{equation}
Dieses $\phi(x)$ ist tatsächlich eine bei $x$ lokalisierte Observable, denn es gilt:
\begin{equation}
   [\phi(x),\phi(y)] = 0  \quad \text{ falls $x$ raumartig zu $y$ liegt }.
\end{equation}
Wir interpretieren $\phi(x)$ daher als eine physikalische Messgröße am Punkt $x$ (auch wenn wir den Bauplan des zugehörigen ,,Messapparats`` hier nicht angeben -- uns interessieren nur die abstrakten Eigenschaften von $\phi$).

In der wechselwirkenden Situation versuchen wir, den Ausdruck \eqref{eq:phidef} zu verallgemeinern:
\begin{equation}
  \phi(x) := \int d\theta \, \Big(e^{ip(\theta) \cdot x} {\zd}(\theta) + e^{-ip(\theta) \cdot x} {z}(\theta) \Big).
\end{equation}
Man findet jedoch, dass dieser Operator \emph{nicht} lokal bei $x$ ist:
\begin{equation}\label{eq:phinotlocal}
   [\phi(x),\phi(y)] {\neq} 0  \quad \text{ selbst wenn $x$ raumartig zu $y$ liegt. }
\end{equation}
Welche Bedeutung können wir $\phi(x)$ also geben? Wir definieren ein zweites Quantenfeld,
\begin{equation}
  \phi'(x) := U(j) \phi(-x) U(j).
\end{equation}
Es stellt sich dann heraus, dass
\begin{equation}
   [\phi(x),\phi'(y)] = 0  \quad \text{ falls $x$ raumartig \emph{links von} $y$ liegt}.
\end{equation}
(Die Aufteilung des raumartigen Gebiets in ,,links`` und ,,rechts`` ist eine spezielle Eigenschaft der 1+1-dimensionalen Raumzeit.) Wir interpretieren dies als eine Lokalisierung von $\phi(x)$ in einem nach links geöffneten Keilgebiet $\wcal_x'$ und von $\phi'(y)$ in dem nach rechts geöffneten Keilgebiet $\wcal_y$ (Abb.~\ref{fig:wedges}).
 
Das Feld $\phi(x)$ ist also in einem unendlich ausgedehnten Keilgebiet lokalisiert. Für die Interpretation im Sinne der Quantenfeldtheorie, z.B. im Rahmen von Streutheorie, benötigt man jedoch Observablen in \emph{beschränkten} Gebieten (wenn auch nicht notwendigerweise an Raum-Zeit-Punkten). 

Gibt es solche Observablen in beschränkten Gebieten $\ocal$ in unseren Modellen? Nach \eqref{eq:phinotlocal} gehören die Felder $\phi(x)$, $\phi'(x)$ nicht dazu (außer für den Fall des freien Feldes, $S=1$). Auch Polynome in diesem Feldern erfüllen die benötigten Vertauschungsrelationen \eqref{eq:ABlocal} nicht (mit Ausnahme bestimmter Polynome im Fall $S=-1$, siehe Abschnitt~\ref{sec:even} unten). Erst wenn man zu Grenzwerten solcher Polynome übergeht, also gewissermaßen Potenzreihen im Feld $\phi$ bildet, kann man in $\ocal$ lokalisierte Observablen erhalten.

\begin{figure}
\begin{center}
 \includegraphics[width=0.5\textwidth]{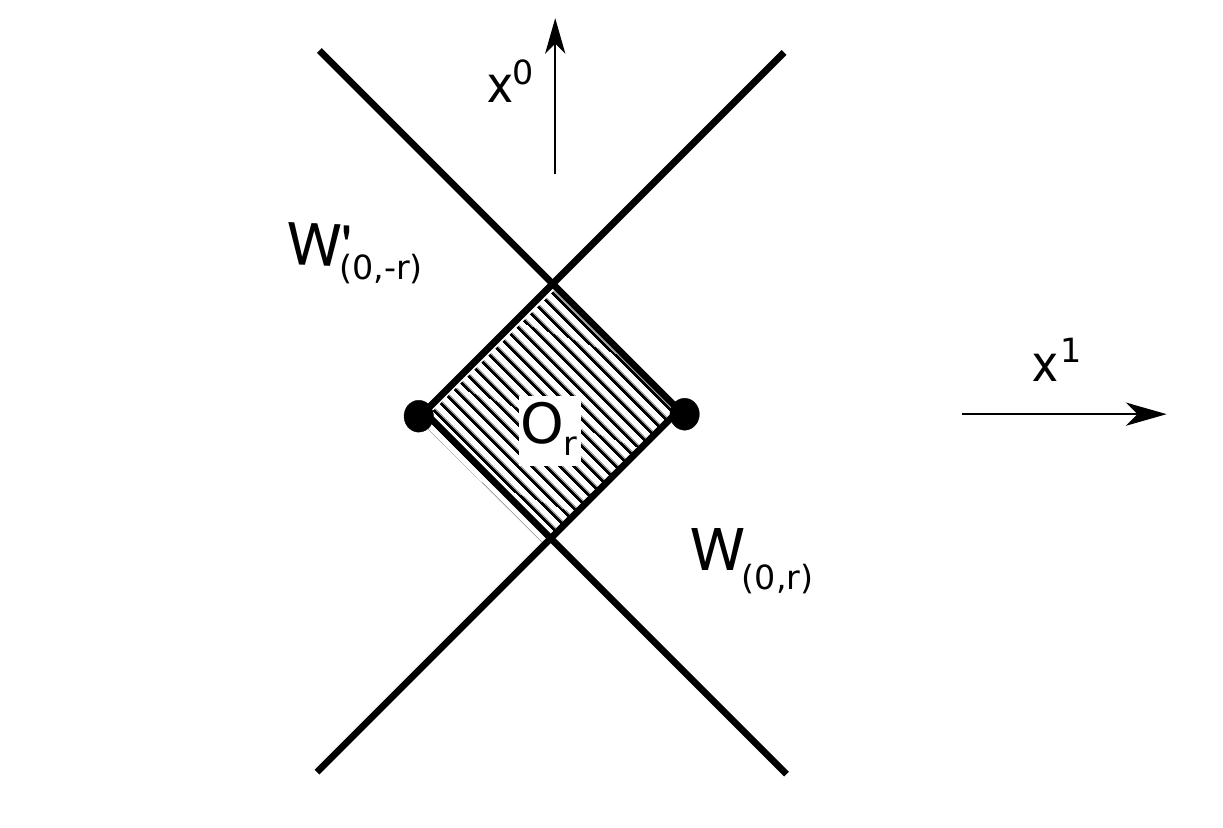}
 \caption{Der Doppelkegel $\ocal_r$ als Durchschnitt zweier Keilgebiete}\label{fig:doublecone}
\end{center}
\end{figure}

Die mathematische Behandlung dieser Grenzwerte ist im Detail sehr aufwändig und erfolgt eher indirekt; wir wollen sie nur sehr grob skizzieren. Aus technischen Gründen geht man zunächst von den (unbeschränkten) Feldern $\phi,\phi'$ zu beschränkten Operatoren in den Keilgebieten über, indem man (unitäre) Exponentialfunktionen der (selbstadjungierten) verschmierten Felder $\phi(f)=\int \phi(x) f(x)dx$ betrachtet. Man definiert die Mengen von Observablen $\A( \mathcal{W}_y )$, $\A( \mathcal{W}_x' )$ dann als von-Neumann-Algebren,\footnote{%
Die Notation $\A'$ steht hierbei für die Kommutante einer Menge $\A$ von Operatoren: $\A' = \{ B : [A,B]=0 \text{ für alle } A \in \A\}$.
} 
\begin{align}
       \A( \mathcal{W}_x' ) &= \{ \exp i \phi(f) \,|\, \supp f \subset \mathcal{W}_x' \} '', \\
       \A( \mathcal{W}_y ) &= \{ \exp i \phi'(f) \,|\, \supp f \subset \mathcal{W}_y \}''.
\end{align}
Beschränkte Gebiete stellt man dann durch Durchschnitte von Keilen dar; insbesondere betrachtet man den Standard-Doppelkegel mir Radius $r$ um den Ursprung,  $\mathcal{O}_r = \mathcal{W}_{(0,-r)} \cap\mathcal{W}_{(0,r)}'$ (Abb.~\ref{fig:doublecone}). Die zugehörige Algebra von Observablen wird dann einfach definiert als
\begin{equation}
    \A(\mathcal{O}_r)  := \A(\mathcal{W}_{(0,-r)}) \cap \A(\mathcal{W}_{(0,r)}').
\end{equation}
Dies ergibt eine konsistente relativistische Quantentheorie, die allen Standard-Axiomen der Quantenfeldtheorie genügt (auch denen, die wir hier nicht angeführt haben). Tatsächlich lassen sich diese Axiome vergleichsweise einfach nachrechnen \cite{Lechner:2008}. 

Die wirklich schwierige Frage in diesem Zusammenhang ist, ob die Algebren $\A(\mathcal{O}_r)$  \emph{nichttrivial} sind, d.h., ob sie irgendeinen Operator außer Vielfachen des Einsoperators enthalten. Tatsächlich gelang es Lechner \cite{Lechner:2008} zu zeigen, dass für eine große Klasse von Streufunktionen $S$ die lokalen Algebren $\A(\mathcal{O}_r)$ sogar sehr groß sind. (Technisch gesprochen ist der Vektor $\Omega$ zyklisch für die Algebren $\A(\mathcal{O}_r)$, d.h., jeder beliebige Vektor im Hilbertraum kann approximiert werden, indem man lokale Operatoren zu einem fixen Gebiet auf das Vakuum anwendet.) 

Lechners Konstruktion verwendet sehr abstrakte mathematische Techniken (u.a.{} die modulare Theorie von Tomita-Takesaki für von-Neumann-Algebren), und die ,,Konstruktion`` lokaler Observabler benötigt letztlich das Auswahlaxiom. Man erhält also keine Aussage über die konkrete Form der Observablen in $\A(\mathcal{O}_r)$, sondern nur über deren abstrakte Existenz. Dies ist dennoch ausreichend, um die Streuzustände und die Streumatrix des Modells vollständig zu bestimmen; es stellt sich heraus, dass unsere Modelle tatsächlich eine faktorisierende Streumatrix mit Zweiteilchen-Streufunktion $S$ besitzen. 

Trotzdem bleibt die Frage, wie man die Operatoren in  $\A(\mathcal{O}_r)$ expliziter beschreiben kann, um damit einen direkteren Zugang zu den lokalen Observablen und ihren Eigenschaften zu erhalten. Wir werden versuchen, darauf eine Antwort zu geben.

\section{Operatorentwicklung und Lokalität}

Um lokale Operatoren $A$ genauer untersuchen und charakterisieren zu können, werden wir eine bestimmte Reihenentwicklung verwenden. Wir formulieren diese zu\-nächst wieder für das freie Feld. Es ist bekannt (siehe z.B. \cite[Sec.~6]{Ara:lattice} oder \cite[Sec.~4.2]{Wei:qft1}), dass sich \emph{jeder} Operator im Hilbertraum des freien Feldes (unabhängig von Lokalitätseigenschaften) folgendermaßen in eine Reihe von normalgeordneten Erzeugern und Vernichtern entwickeln lässt:
\begin{equation}\label{eq:expfree}
 A = \sum_{m,n=0}^{\infty} \int \frac{d^m\theta d^n\eta}{m!n!} f_{mn}(\boldsymbol{\theta},\boldsymbol{\eta}) 
a^{\dagger}(\theta_1) \ldots a^{\dagger}(\theta_m) a(\eta_1) \ldots a(\eta_n).
\end{equation}
Die Koeffizientenfunktionen $f_{mn}$ lassen sich dabei explizit aus $A$ berechnen:
\begin{equation}\label{eq:fmnfree}
 f_{mn}(\boldsymbol{\theta},\boldsymbol{\eta}) 
  = \big(\Omega, [a(\theta_1),\ldots,[a(\theta_m), [a^\dagger(\eta_1),\ldots,[a^\dagger(\eta_n),\, A \, ] \ldots ] \Omega\big).
\end{equation}
Welche Eigenschaften haben diese Koeffizienten $f_{mn}$ nun, wenn $A$ im Standard-Doppelkegel $\ocal_r$ lokalisiert ist? Im Wesentlichen sieht man dies wie folgt. Man schreibt die Erzeuger und Vernichter in \eqref{eq:fmnfree} als Linearkombinationen der Zeit-0-Felder $\varphi, \pi$ bzw.~deren Fourier-Transformierter. Wegen der Lokalität von $A$ verschwinden die Kommutatoren von $A$ mit $\varphi(x^1), \pi(x^1)$, wenn $|x^1|$ genügend groß ist, nämlich so groß, dass $(0,x^1)$ raumartig zu $\ocal_r$ liegt. Daher sind die Funktionen $f_{mn}$ im Wesentlichen die Fourier-Transformierten von Funktionen mit kompaktem Träger, und als solche sind sie ganz analytisch. Aufgrund der Struktur von \eqref{eq:fmnfree} haben diese analytischen Funktionen weitere Symmetrieeigenschaften.

Im Einzelnen ergibt sich folgendes. Ist der Operator $A$ in $\mathcal{O}_r$ lokalisiert, dann gibt es analytische Funktionen $F_k:\mathbb{C}^k \to \mathbb{C}$, so dass
\begin{equation}
   f_{mn}(\thetav, \etav) = F_{m+n}(\theta_1,\ldots,\theta_m, \eta_1+i\pi, \ldots, \eta_n+i\pi).
\end{equation}
Diese $F_k$ haben folgende Eigenschaften:
\begin{enumerate}[(i)]
 \item Sie sind symmetrisch in ihren Argumenten (als Konsequenz der Vertauschungsrelationen \eqref{eq:ccr}),
 \item Sie sind  {$2 i \pi $-periodisch} in jedem Argument (da $p(\theta)$ diese Periodizität besitzt),
 \item Sie erfüllen gewisse $r$-abhängige Schranken in der mehrdimensionalen komplexen Ebene (ähnlich wie im Satz von Paley-Wiener).
\end{enumerate}
Diese Bedingungen sind in einem bestimmten Kontext\footnote{%
Z.B. als Spezialfall von \cite[Theorem~5.4]{Cadamuro:2012} für triviale Streumatrix $S=1$}
auch hinreichend für die Lokalität von $A$ in $\ocal_r$, d.h., man erhält eine vollständige Charakterisierung der lokalen Observablen durch ihre Koeffizientenfunktionen $f_{mn}$.

Wir versuchen nun, diese Methode für allgemeine Streufunktionen $S$ auszudehnen. Die offensichtlich Idee ist, in \eqref{eq:expfree} die Erzeuger und Vernichter $\ad(\theta),a(\theta)$ durch ,,wechselwirkende`` Objekte $\zd(\theta),z(\theta)$ zu ersetzen. Tatsächlich kann man zeigen \cite{BostelmannCadamuro:expansion}, dass sich jeder Operator $A$ im jeweiligen Hilbertraum schreiben lässt als
\begin{equation}\label{eq:expinter}
 A = \sum_{m,n=0}^{\infty} \int \frac{d^m\theta d^n\eta}{m!n!} f_{mn}(\boldsymbol{\theta},\boldsymbol{\eta}) 
{z^{\dagger}}(\theta_1) \ldots {z^{\dagger}}(\theta_m) {z}(\eta_1) \ldots {z}(\eta_n) .
\end{equation}
Wiederum lassen sich die Koeffizientenfunktionen $f_{mn}$ explizit als Funktion von $A$ schreiben, jedoch ist dieser Ausdruck wesentlich komplizierter als \eqref{eq:fmnfree}. 

Was geschieht, wenn $A$ im Doppelkegel $\ocal_r$ lokalisiert ist? Dies ist weit schwieriger zu beantworten als für das freie Feld. Es ist sinnvoll, zunächst Operatoren zu betrachten, die im Keil $\wcal_0'$ lokalisiert sind. Für diese erhält man wiederum eine analytische Fortsetzung der Funktionen $f_{mn}$, jedoch (wegen der weniger strengen Lokalisierung) nicht auf die gesamte mehrdimensionale komplexe Ebene. Der Doppelkegel $\ocal_r$ ist nun der Durchschnitt zweier Keilgebiete, und entsprechend erhält man für $A\in\A(\ocal_r)$ \emph{zwei} Sätze von analytischen Funktionen $F_k$. Diese lassen sich dann zu je einer Funktion $F_k$ zusammenfügen; man erhält so Funktionen auf ganz $\cbb^k$, die jedoch in der Regel nicht mehr ganz analytisch sind, sondern nur meromorph (d.h., sie können Pole endlicher Ordnung besitzen). Ein wesentlicher Punkt in der Konstruktion \cite{Cadamuro:2012}, die wir hier nicht explizit angeben, ist, dass der eine beteiligte Keil gegenüber dem anderen am Koordinatenursprung gespiegelt ist -- es ist daher wichtig, zu wissen, wie Raum-Zeit-Spiegelungen auf die Koeffizienten $f_{mn}$ wirken \cite[Prop.~3.11]{BostelmannCadamuro:expansion}.

Insgesamt ergibt die Konstruktion folgendes: Ist der Operator $A$ lokal in $\mathcal{O}_r$, dann gibt es meromorphe Funktionen $F_k:\mathbb{C}^k \to \mathbb{C}$ so dass
\begin{equation} 
   f_{mn}(\thetav, \etav) = F_{m+n}(\theta_1+i0,\ldots,\theta_m+i0, \eta_1+i\pi-i0, \ldots, \eta_n+i\pi-i0).
\end{equation}
Die $F_k$ haben die folgenden Eigenschaften (vgl.~\cite{SchroerWiesbrock:2000-1}):
\begin{enumerate}[(i)]
 \item Sie sind $S$-symmetrisch: 
\begin{equation}
  F_k(\ldots \zeta_j,\zeta_{j+1},\ldots) = S(\zeta_{j+1}-\zeta_j) F_k(\ldots \zeta_{j+1},\zeta_j,\ldots).
\end{equation}

\item Sie sind $S$-periodisch:  
\begin{equation}
F_k(\ldots, \zeta_j+2 i \pi, \ldots) =
\Big(\prod_{\substack{i=1 \\ i \neq j}}^k S(\zeta_i-\zeta_j)\Big)
F_k(\ldots, \zeta_j,\ldots). 
\end{equation}

\item Sie haben Pole bei $\zeta_n-\zeta_m = i \pi$ ($m<n$) mit Residuen
\begin{equation}\label{eq:recursion}
\operatorname{Res}_{\zeta_n-\zeta_m = i \pi} F_k(\zetav) = 
= - \frac{1}{2\pi i }
\Big(\prod_{j=m}^{n} S(\zeta_j-\zeta_m) \Big)
\Big(1-\prod_{p=1}^{k} S(\zeta_m-\zeta_p) \Big)
F_{k-2}( \hat\zetav ).
\end{equation}
(Hierbei  entsteht $\hat\zetav\in\cbb^{k-2}$ aus $\zetav\in\cbb^k$, indem man die Komponenten $\zeta_m$ und $\zeta_n$ weg lässt.)
 \item Sie erfüllen gewisse $r$-abhängige Schranken (ähnlich denen im Satz von Paley-Wiener, aber leicht modifiziert).
\end{enumerate}
Wiederum sind diese Bedingungen in einem gewissen Kontext auch hinreichend \cite{Cadamuro:2012,BostelmannCadamuro:characterization}, man erhält also eine vollständige Charakterisierung der lokalen Observablen.

\section{Beispiele lokaler Operatoren} \label{sec:ops}

Nachdem wir nun die lokalen Observablen \emph{charakterisiert} haben, wollen wir diese Information verwenden, um konkrete lokale Observable zu \emph{konstruieren}. Das heißt, wir wollen Beispiele von meromorphen Funktionen $F_k$ angeben, die die oben genannten Bedingungen erfüllen; der durch \eqref{eq:expinter} gegebene Operator $A$ ist dann automatisch lokal. 

Hierbei müssen wir zwei Schwierigkeiten unterschiedlicher Natur überwinden:
\begin{enumerate}[(a)]
   \item ein \emph{kombinatorisches Problem}: Wie findet man eine Folge von Funktionen $F_k$ mit der gewünschten Symmetrie,  Periodizität und Residuenstruktur?
   \item ein \emph{Konvergenzproblem}: Definiert die Reihe \eqref{eq:expinter} tatsächlich einen \emph{Operator} im mathematischen Sinn, oder nur eine quadratische Form?
\end{enumerate}
Wichtig hierbei ist, dass die Summe über $m,n$ in \eqref{eq:expinter} in der Regel unendlich ist, denn die Folge der $F_k$ kann wegen der Residuenbedingung \eqref{eq:recursion} nicht abbrechen, außer für sehr spezielle Wahl von $S$ (siehe Abschnitt~\ref{sec:even} unten).

Der Aspekt (a) scheint für allgemeines $S$ zwar im Prinzip lösbar (siehe \cite{FringMussardoSimonetti:1993} für den Fall des sinh-Gordon-Modells), ist aber im Detail recht aufwändig. Aspekt (b) ist im allgemeinen Fall bisher nicht verstanden. Wir beschränken uns deshalb hier auf einen einfachen, aber nicht uninteressanten Spezialfall: wir wählen die Streufunktion $S=-1$ konstant.

Dies ist die quantenfeldtheoretische Version des sogenannten massiven Ising-Modells, das sonst aus der statistischen Physik bekannt ist. (Genauer hängt es mit dem Kontinuumlimes des Ising-Modells oberhalb der kritischen Temperatur zusammen.) Der Spezialfall $S=-1$ vereinfacht die Fragestellung deutlich, denn die Residuen \eqref{eq:recursion} werden nun zu
\begin{equation}\label{eq:ising-recursion}
 \operatorname{Res}_{\zeta_n-\zeta_m = i \pi} F_k(\zetav) 
=  \frac{1}{2\pi i } (-1)^{m+n} (1-(-1)^{k}) F_{{k-2}}(\hat\zetav).
\end{equation}
Da diese Gleichung eine Beziehung zwischen $F_k$ und $F_{k-2}$ herstellt, bietet es sich an, die Fälle von geradem und ungeradem $k$ getrennt voneinander zu betrachten.

\subsection{Ising-Modell, gerade Operatoren} \label{sec:even}

Wir betrachten zunächst gerade Werte von $k$. In diesem Fall verschwinden die Residuen der $F_k$, da die rechte Seite von \eqref{eq:ising-recursion} zu Null wird. Es ist dann sehr einfach, Beispiele von lokalen Operatoren zu finden (siehe auch \cite{BuchholzSummers:2007}). 
Wir setzen etwa
\begin{equation}\label{eq:f2ising}
  F_{2}(\zeta_{1},\zeta_{2})=\sinh \Big(\frac{\zeta_{1}-\zeta_{2}}{2} \Big)\tilde{g}(\mu E({\zetav})),
\end{equation}
wobei
\begin{equation}
 E(\zetav) := \sum_{j=1}^k \cosh \zeta_j
\end{equation}
und wobei $g$ eine glatte Funktion mit Träger im Intervall $[-r,r]$ ist. Alle anderen $F_k$ ($k \neq 2$) setzen wir zu Null. Es ist relativ einfach nachzurechnen, dass die Lokalisierungsbedingungen an $F_k$ erfüllt sind. Man erhält durch die Reihe \eqref{eq:expinter} -- die in diesem Fall tatsächlich eine endliche Summe ist -- dann eine ,,gute`` lokale Observable 
(genauer, einen abschließbaren Operator $A$, der mit der von-Neumann-Algebra $\A(\ocal_r)$ affiliiert ist.) 

Durch Modifikation von \eqref{eq:f2ising}, etwa durch Multiplikation mit symmetrischen Polynomen in $\exp \zeta_j$, lassen sich leicht weitere Beispiele angeben. Interessanterweise liegt auch die Energiedichte des Ising-Modells in dieser Klasse von Observablen \cite{BostelmannCadamuroFewster:2013}.

\subsection{Ising-Modell, ungerade Operatoren} \label{sec:odd}

Für ungerades $k$ ist das kombinatorische Problem etwas schwieriger. In der Residuenbedingung \eqref{eq:ising-recursion} ist der Vorfaktor auf der rechten Seite nun nicht Null, allerdings konstant. Die Funktionen $F_k$ sind zumindest heuristisch bekannt \cite{SchroerTruong:1978}. Wir setzen
\begin{equation}
  F_{2k+1} (\zetav) =  \frac{1}{(4 \pi i)^{k} k!} \tilde g( \mu E(\zetav) ) \sum_{\sigma \in \mathfrak{S}_{2k+1}} \operatorname{sign} \sigma \prod_{j=1}^k \tanh \frac{ \zeta_{\sigma(2j-1)} - \zeta_{\sigma(2j)}}{2},
\end{equation}
wobei wiederum $g$ eine glatte Funktion mit Träger im Intervall $[-r,r]$ ist, und $\mathfrak{S}_{2k+1}$ die Gruppe der Permutationen von $2k+1$ Elementen. Alle Funktionen $F_{2k}$ setzen wir zu Null.

Aufgrund der Residuenstruktur der $\tanh$-Funktion erfüllen diese Funktionen alle oben genannten Bedingungen (einschließlich der Paley-Wiener-ähnlichen Schranken, auf die wir hier nicht näher eingegangen sind). 

Es ist nun ein schwieriges Problem, die Konvergenz der Reihe in \eqref{eq:expinter} als Operator auf einem geeignet großen Definitionsbereich zu kontrollieren. Dies gelingt tatsächlich, falls die Funktion $g$ in einer geeigneten Jaffe-Klasse liegt \cite[Sec.~9]{Cadamuro:2012}. Wie zuvor erhält man einen abgeschlossenen, mit der von-Neumann-Algebra  $\A(\ocal_r)$ affiliierten Operator.

\section{Fazit}

Quantenfeldtheorien mit faktorisierender Streumatrix in 1+1 Dimensionen lassen sich mathematisch rigoros konstruieren. Dies geschieht direkt im relativistischen Kontext, ohne Verwendung einer klassischen Lagrange-Funktion, eines Gitterlimes oder einer Wick-Rotation. 

Die Lokalität von Observablen kann durch die Analytizitätseigenschaften ihrer Entwicklungskoeffizienten in einer Reihenentwicklung charakterisiert werden. Zumindest in einfachen Fällen (Ising-Modell, $S=-1$) kann man dies zur expliziten Konstruktion lokaler Observabler als abschließbare Operatoren verwenden. Dies ist ein wesentlicher Fortschritt gegenüber früheren Ansätzen, insbesondere dem des Formfaktor-Programms \cite{BabujianFoersterKarowski:2006}. Dieses basiert auf einer -- von \eqref{eq:expinter} etwas verschiedenen -- Reihenentwicklung lokaler Punktfelder, und man versucht, deren Konvergenz in Wightman-$n$-Punkt-Funktionen zu kontrollieren, was aber nicht vollständig gelingt. Hierbei soll angemerkt werden, dass die Resultate in Abschnitt~\ref{sec:ops} \emph{keine} $n$-Punkt-Funktionen liefern, da die konstruierten Operatoren $A$ eventuell keinen invarianten dichten Definitionsbereich besitzen und ihre Produkte daher möglicherweise nicht existieren. Dies ist auch konzeptionell nicht notwendig; die Affiliiertheit des Operators zu den lokalen Algebren $\A(\ocal_r)$ reicht für eine Interpretation aus.

Es wäre nun interessant, Beispiele für allgemeineres $S$ zu konstruieren. Zumindest in bestimmten Modellen, wie etwa sinh-Gordon \cite{FringMussardoSimonetti:1993}, sind die Lösungen im Prinzip bekannt, allerdings ist die Konvergenz der Reihe \eqref{eq:expinter} ein offenes Problem. Eine Verallgemeinerung der Methoden aus \cite{Cadamuro:2012} könnte dieses Konvergenzproblem jedoch lösen.

Wir haben uns hier auf Modelle in 1+1 Raum-Zeit-Dimensionen beschränkt. Natürlich stellt sich die Frage, ob die gleichen Techniken auch wechselwirkende Modelle in 3+1-dimensionaler Raum-Zeit liefern können. Dies ist so direkt wohl nicht der Fall: Die stark vereinfachte Situation der faktorisierenden Streumatrix, die keine Teilchenerzeugungsprozese zulässt, ist zu simpel, um Streuung in physikalischer Raumzeit zu beschreiben. Man kann mit einer analogen Konstruktion Operatoren erhalten, die im Durchschnitt zweier höherdimensionaler Keilgebiete lokalisiert sind \cite{BuchholzSummers:2007} -- allerdings ist dieser Durchschnitt in 3+1 Dimensionen nicht beschränkt, sondern unendlich ausgedehnt. Die Existenz lokaler Operatoren in beschränkten Gebieten ist in diesem Kontext nicht zu erwarten. Allerdings könnte eine Variante der oben beschriebenen Charakterisierung lokaler Operatoren verwendet werden, um die Abwesenheit lokaler Operatoren im Sinne eines ,,No-go-Theorems`` streng zu beweisen.

\bibliographystyle{germanunsrt}
\bibliography{integrable}

\end{document}